\documentclass[prb,preprint,showpacs,endfloats,amsmath,amssymb]{revtex4-1}

\usepackage{graphicx}
\usepackage{color}
\usepackage{textcomp}

\newcommand{\jc}{$j_\mathrm{c}$}
\newcommand{\Bs}{$B^*$}
\newcommand{\Tc}{$T_\mathrm{c}$}

\newcommand{\extder}{\text{d}} 
\newcommand{\half}{\tfrac{1}{2}}
\newcommand{\quarter}{\tfrac{1}{4}}
\newcommand{\abs}[1]{\left\lvert#1\right\rvert}
\newcommand{\norm}[1]{\lVert#1\rVert}
\DeclareMathOperator{\curl}{\nabla \times}

\begin{document}

\title[]{Modelling reduced field dependence of \jc\ in YBCO films with nanorods}

\author{H. Palonen$^{1,2}$}
\email[To whom correspondence should be addressed: ]
{heikki.palonen@utu.fi}
\author{J. J\"aykk\"a$^3$}
\author{P. Paturi$^1$}

\affiliation{$^1$Wihuri Physical Laboratory, Department of Physics and 
Astronomy, FIN-20014 University of Turku, Finland} 
\affiliation{$^2$Graduate School of Materials Research, Turku, Finland}
\affiliation{$^3$School of Mathematics, University of Leeds, LS2 9JT, 
United Kingdom}

\date{\today}

\begin{abstract}

Flux pinning is still the main limiting factor for the critical current of 
the high-temperature superconductors in high fields. In this paper we 
model the field dependence of the critical current in thin films with 
columnar defects aligned with the field. The characteristic shape of the 
critical current of the BaZrO$_3$ doped YBa$_2$Cu$_3$O$_{6+x}$ thin films 
is reproduced and explained. The model is based on solving the Ginzburg-
Landau equations with columnar defects present in the lattice. The size of 
the columnar defect is found to be of key importance in explaining the 
rounded shape of the critical current of the BaZrO$_3$ doped 
YBa$_2$Cu$_3$O$_{6+x}$ thin films. It is also found that the size of the 
rod changes the long range order of the vortex lattice.

\end{abstract}


\maketitle

\section{Introduction}

The critical current density, \jc, of high-temperature superconductors
in magnetic field is mainly limited by flux pinning \cite{Foltyn4}. Most of
the envisioned applications, such as generators and fault current
limiters, require the superconductors to have high critical current at
magnetic fields of 3-5 T. In this range \jc\ is proportional
to the fraction of vortices pinned in different pinning sites
\cite{Pan3}. Each individual pinning site provides a pinning force,
$f_\mathrm{p}$, which depends on the type of the pinning
site. Typical pinning sites in YBa$_2$Cu$_3$O$_{6+x}$ (YBCO)
superconductors are dislocations, twins, antiphase boundaries,
impurities and grain boundaries \cite{Foltyn4}. Out of these dislocations
have been found most effective \cite{Dam2}, since they have the same shape
as the vortices. Their effectiveness is limited by the small size of
the core of the dislocation, 0.3 nm \cite{Svetchnikov1}, which is only
a fraction of the vortex size in YBCO ($\xi = 1.5$ nm at 0 K
\cite{Blatter4}). Twin planes pin vortices effectively if the current is 
parallel to the twin plane and thus the Lorentz force perpendicular to the 
plane, unfortunately if the current is perpendicular to the twin plane, the 
plane channels the vortices for easy movement, and therefore the \jc\ 
decreases \cite{Blatter4,Crabtree1}.

Doping the YBCO with e.g.\ BaZrO$_3$ (BZO) or BaSnO$_3$ (BSO) leads to
formation of nonsuperconducting nanorods into the superconducting
matrix \cite{MacManusDriscoll2,MacManusDriscoll8,Varanasi2,Varanasi5}.
The nanorods act as very efficient pinning sites, since their diameter
is around 5--10 nm \cite{Peurla3,Augieri5} and therefore around the same 
size as the vortex core. The nanorods are $c$-axis oriented and increase
\jc\ at high fields especially when $B\parallel c$ \cite{Peurla3,Paturi15}. 


The shape of the \jc$(B)$ is most often described with the accommodation
field, \Bs, where the low field plateau ends and the exponent $\alpha$,
which describes the decrease of \jc\ above \Bs\ with field $B^{-\alpha}$
\cite{Goyal4,Peurla3,Paturi17,Dam2,Klaassen2}. For typical undoped YBCO
thin films the \Bs $=$ 40--100 mT and $\alpha \approx 0.6$ \cite{Klaassen2}.
In BZO-doped samples the \Bs\ increases up to 0.5 T and $\alpha$ decreases
to 0.2 -- 0.4 \cite{Peurla3}. The $\alpha$-value observed in undoped films
is predicted by theories of strong sparse pinning sites
\cite{Beek1,Nelson4}, but the lower value in the doped films has not been 
predicted, and finding a simple explanation for the lowered $\alpha$ value 
seems to be difficult due to the vortex-vortex interactions involved.

The problem in modelling flux pinning is that all the real samples
contain many different kinds of pinning sites and the resulting
\jc$(B)$ consists of all the different interactions. It is
relatively simple to calculate the pinning force of a single type of
pinning site at low field \cite{Blatter4}, but even increasing the
field, when vortex-vortex interactions come into play, makes the
calculations in closed form  impossible. Statistical approach
has been used e.g.\ in refs. \onlinecite{Pan3,Paturi16,Paturi18,Long4,Long5}
and these models describe usually the shape of the experimental
curves very well, but the understanding of e.g.\ the vortex paths
inside the superconductors is limited.

The Ginzburg-Landau equations, although originally meant to be used
close to the phase transition, describe superconductors well at lower 
temperatures too. Solving the minimum energy configuration
for a certain set of parameters gives us the spatial variation of the
order parameter $\psi$, which can be used to calculate the fraction of
the pinned vortices as a function of e.g.\ the magnetic field or the 
density, shape and size of the pinning sites. The Ginzburg-Landau equations 
have been solved for small particles \cite{BarbaOrtega1} and with one or 
several pinning sites using $\delta T_\mathrm{c}$ pinning \cite{Machida1} 
and by restricting the order parameter \cite{Nakai1}. However, Crabtree
\textit{et al.} \cite{Crabtree1} present a large scale model with flux
pinning that is similar to this work. Unfortunately, they do it for
a relatively low $\kappa = 4$ (low for modelling YBCO) and concentrate
on the dynamics of the vortex trajectories instead of the critical
current that is the focus of this work. We present a method for
doing large scale computation of pinning in superconductors that are
close to realistic size and calculate the field depedence of the critical 
current. In this paper we consider the case where magnetic field is
parallel to the columnar defects. The results are directly compared to data 
from thin film YBCO samples.




\section{Computational model}

\subsection{Solving the Ginzburg-Landau equations}

We chose to solve the static Ginzburg-Landau equations by finding a
(local) minimum of the associated energy functional. For computational
purposes, we write the energy in a dimensionless form. The only
dimensional value is the overall energy scale, which does not affect
the solutions and the dimensionless energy is
\begin{align}
  \label{eq:scaled_E}
  \begin{split}
    E = 
    \int \extder^3 {x}
    \bigl(
        \half\norm{(\nabla + i {\vec{A}}){\psi}}^2 +
        \half\norm{\curl{{\vec{A}}}}^2\\
        + \quarter \kappa^2 (\abs{\psi}^2-1)^2
    \bigr),
  \end{split}  
\end{align}
where $\kappa = \sqrt{\beta/(2\mu_0\hbar^2\gamma^2q^2)}$
is the dimensionless Ginzburg-Landau parameter and
$\gamma = 1/(4m_\mathrm{e})$ and $q = 2e$, the penetration depth is 
absorbed in the overall energy scale and therefore $\lambda = 1$. 
Naturally, the coherence length is now simply $\xi = 1/\kappa$. Equation 
\eqref{eq:scaled_E} is then discretised as described in ref.\
\onlinecite{Jaykka2}. In short, this discretization preserves the gauge 
invariance of the system and allows us to solve the equations without 
choosing and enforcing a gauge and the associated problems.

The solutions are then found using the TAO \cite{tao-user-ref} and
PETSc \cite{petsc-web-page,petsc-user-ref,petsc-efficient} massively
parallel numerical libraries. We account for the impurities in the
physical system by using a bound constrained variant of the limited
memory quasi-Newton algorithm (also called a variable metric
algorithm) with BFGS \cite{Broyden1, Fletcher1,
Goldfarb1, Shanno1} formula for Hessian
approximations. This is provided by the TAO library. The impurities
are then modelled by setting appropriate constraints for the
Cooper pair density at the impurity locations. The gradients
required by the algorithm are computed from the discretized energy in
a straightforward manner. 

The correctness and accuracy of the program was tested by solving the
equations for a single Abrikosov vortex. At a $102^2$ lattice and
lattice constant of $0.1 \approx 0.07\xi$, the deviation from correct 
total energy is less than $0.02$ \% with most of the deviation resulting 
from the small box: increasing the number of lattice points increases the 
accuracy. A smaller lattice constant does not increase the accuracy as 
much. Note that due to the rescaling, the lattice constant is in units of
penetration depth $\lambda$. To be able to resolve the core of the vortex 
the lattice constant has to be smaller than $\xi$. This means that a 
lattice constant $h$ of approximately $0.1 \xi$ gives very accurate 
results. Computer memory is a limiting factor in the computations. In 
practice, $h \approx 0.3 \xi$ yields accurate enough results and requires
$\approx 97$ \% less computer memory.

At the $y$- and $z$-boundaries of the calculation $\psi$ is set to 
zero and the vector potential is kept fixed to simulate an external 
magnetic field. The $x$-boundary is periodic. Lattice sizes used in 
simulations were typically $500 \times 520 \times 50$ of which 10 
pixels near the boundaries were used as vacuum ($\psi = 0$, $\vec{A}$ 
is free). The vacuum between the sample and the calculation boundary 
allows for the magnetic field to bend around the sample which makes 
the external field of the simulation comparable to the actual field of 
the measurement of a thin film in a magnetometer.

\begin{figure}
  \includegraphics[width=8.6cm]{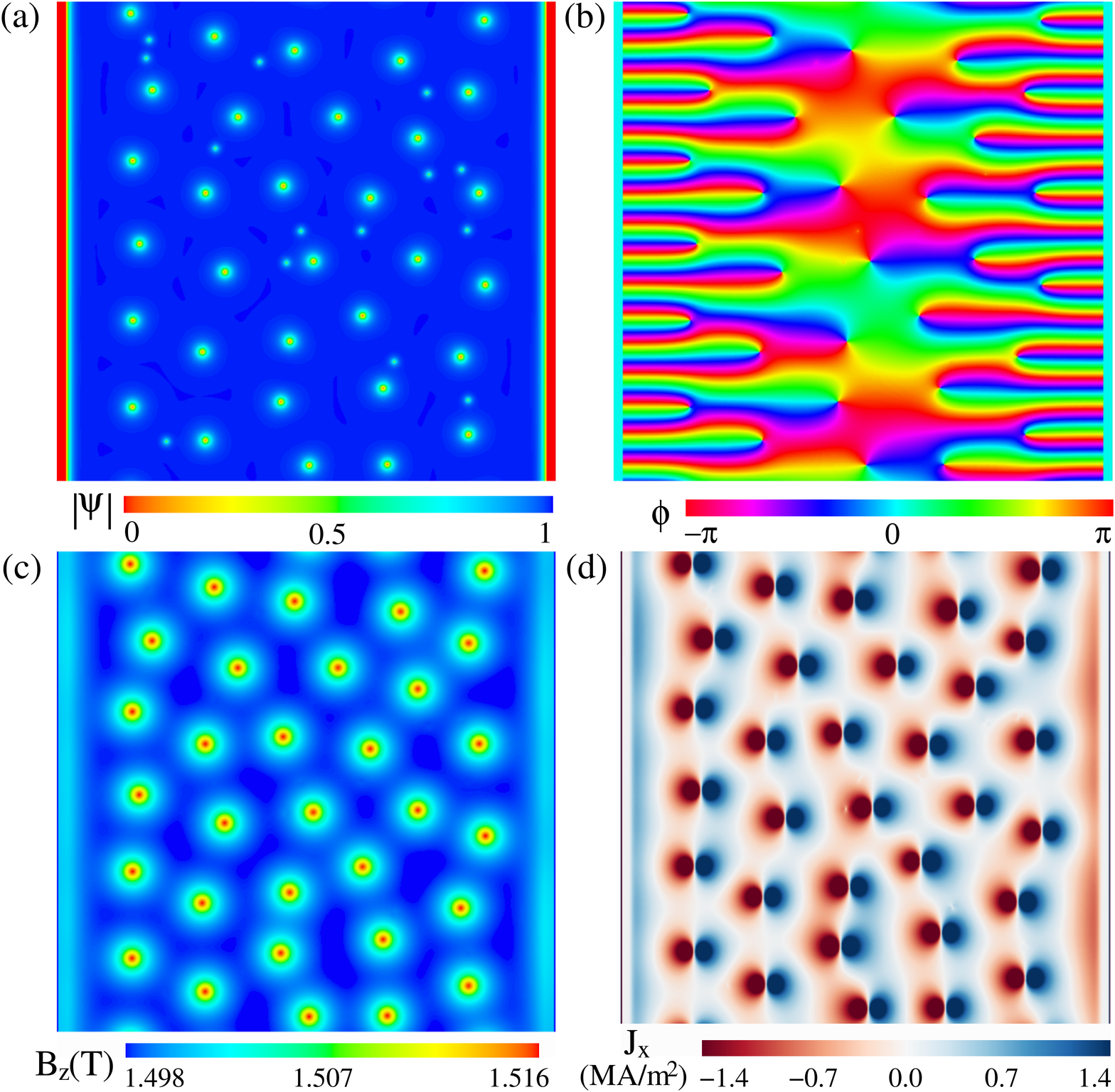}
    \caption{
      a) The absolute value of the calculated order parameter,
      $\abs{\psi}$, shown on a plane sliced perpendicular to
      $c$-axis.
      b) The phase of the order parameter in the same calculation. 
      c) The magnetic field density calculated from the
      vector potential in the same simulation and
      d) the current flowing in the periodic direction (up and down). Note 
      that the scale is cut off short so that also the shielding current is 
      visible.
    }
  \label{simulation}
\end{figure}

As an example of a simulation result fig.\ \ref{simulation} shows the 
absolute value of $\psi$ (fig.\ \ref{simulation}a) and the phase
(fig.\ \ref{simulation}b) with $B= 3$ T in a sample with dislocations. Using 
also the vector potential $\vec{A}$, the magnetic field 
$B_z=(\nabla \times \vec{A})_z$ (fig.\ \ref{simulation}c) and the current in 
the periodic direction $j_x = (\nabla \times \vec{B})_x /\mu_0$
(fig.\ \ref{simulation}d) were calculated. Note that also the shielding 
currents are visible. The calculation always yielded zero current in the 
vacuum, thus confirming the validity of the vacuum spacer layer.

\subsection{Introducing pinning to the model and calculating \jc}

Flux pinning was modelled by locally restricting the maximum possible
value of the real and imaginary parts of $\psi$ which then represents a
pinning site. The chosen maximum value was 0.1 which corresponds to
limiting the maximum value of  $\abs{\psi}$ between 0.1 and
$\sqrt{2}/10$. The limit was not set to zero because that would have 
made the analysis of the result considerably more difficult. With this 
limit vortices can be  defined to be the region of the calculation 
lattice where $\abs{\psi} < 0.1$. 

The pinning sites had the same physical size and shape as is deduced from 
transmission electron microscopy (TEM) data: The BZO-nanorods were modelled 
as randomly distributed sample penetrating rods with a diameter of 5 nm 
\cite{Peurla3,Augieri5}. The dislocations were $c$-axis aligned sample 
penetrating rods with a diameter of 0.3 nm \cite{Svetchnikov1}, which were 
also randomly distributed in the sample.

The critical current was determined as the fraction of the vortex length
trapped in the pinning site to the total length as in 
ref.\ \onlinecite{Pan3}. Thus, we do not get the absolute values, but the 
field dependence of the \jc\ instead, which can be directly compared to the 
experimental data. The total length of vortices was calculated by following 
each vortex. If a point along the vortex is closer than $1.5\xi$ to a 
pinning site it was counted to the pinned section of the vortex. Finally, 
\jc\ is proportional to the fraction of the pinned sections to the total 
vortex length.

\begin{figure}
  \includegraphics[width=8.6cm]{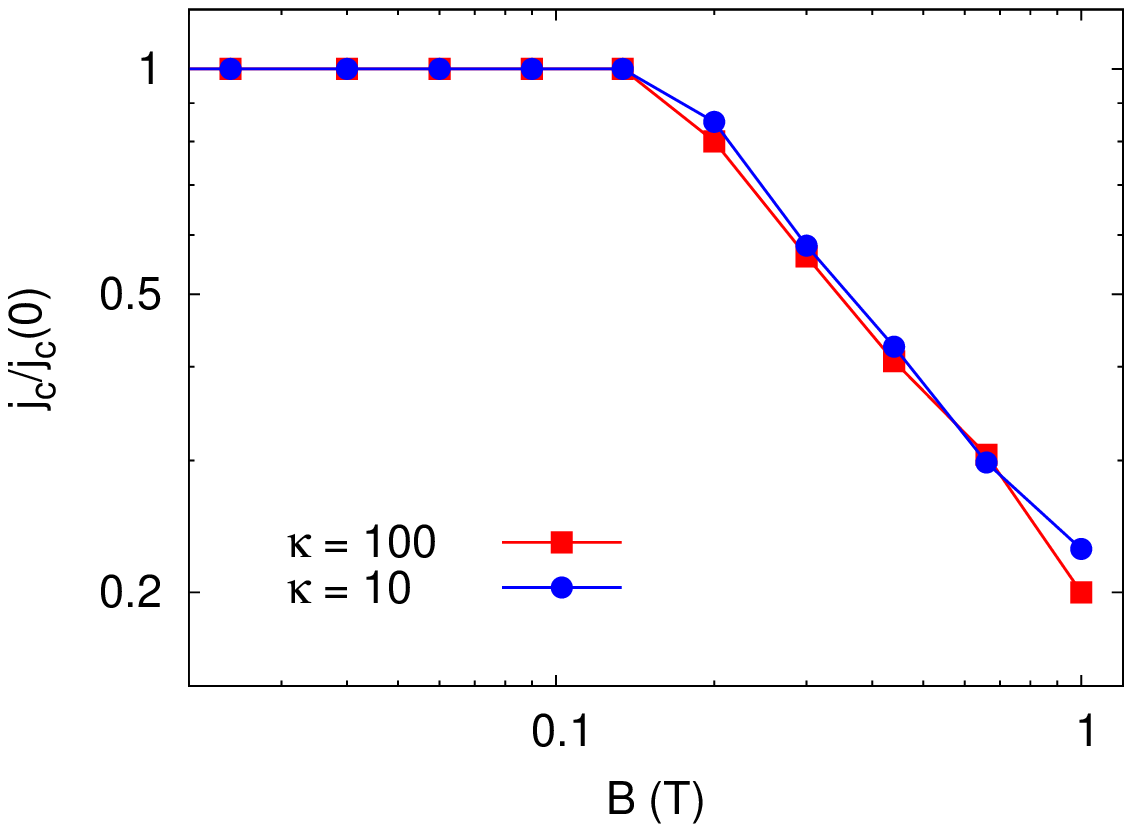}
  \caption{
    The same simulation calculated with $\kappa = 10$ and
    $\kappa=100$. The change in $\kappa$ does not change the
    results.
  } 
  \label{kappa}
\end{figure}

Modelling different pinning site types at the same time requires 
calculations over different length scales which is memory consuming. 
Dislocations are small in size and sparsely distributed which means 
large sample sizes are needed to contain more than a few dislocations. 
As a solution to this memory issue the coherence length and the  
pinscape were scaled up so that $\kappa = 10$. Thus, we can use larger 
lattice constant allowing sparse dislocation densities. The change of
$\kappa$ can be done without major changes to the physics
\cite{Poole2}, since even at $\kappa = 10$, we are at the limit of 
high-$\kappa$ and the magnetic field variation inside the superconductor 
is small. This was also verified by calculating two simulations with 
the same pinscape and resolution in units of $\xi$ but with different
$\kappa$. Figure \ref{kappa} shows the results of such calculations. It 
is easily seen that changing $\kappa$ does not have an appreciable 
effect on the results.

%

\section{Results and discussion}


The field dependence of \jc\ with BZO-nanorods as a pinscape was
simulated for a 230 nm wide sample with thickness of 15 nm that is thick 
enough for the order parameter to reach 1 inside the sample and for all the 
relevant physics since here $\vec{B}$ is always perpendicular to the sample. The 
third direction was periodic with a period of 240 nm. There was also a 5 nm
thick layer of vacuum around the sample. Both the rods and the magnetic
field were parallel to $c$-axis. The magnetic field was varied from 4
T to 0 T with small steps. The result of the calculation with the
previous field value was used as the initial condition for the
calculation with the next field value.  The density of the BZO-nanorods
was set so that the matching fields were $B_\phi = 1$ T 
(30 rods) and $B_\phi = 2$ T (60 rods).

The \jc\ with dislocations as pinning sites was
calculated with a similar calculation grid but its length was scaled 
by a factor of ten, due to the change in $\kappa$, to achieve the 
required low density of dislocations. The fine tuning of the dislocation 
density to match the experimental data was done after the simulation by 
scaling the length of the calculation lattice unit cell. The scaling does 
not alter the results because the 
pinning strength is determined by the ratio $r_\mathrm{r}/\xi$,
and vortex-vortex interactions are affected by the ratio of the average 
pinning site separation to the penetration depth $\lambda$ which is also 
dimensionless. What the scaling does affect are the size of the sample and 
the value of the magnetic field. Thus, the sample size for the dislocation 
simulations was 530 nm (periodic) $\times$ 530 nm $\times$ 34 nm with 
dislocation densities corresponding to matching fields of 90 mT (12 rods), 
180 mT (25 rods) and 360 mT (50 rods). 

\begin{figure}
  \includegraphics[width=8.6cm]{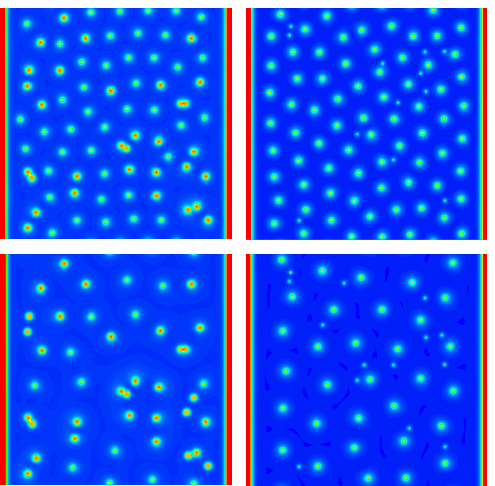}
  \caption{
    The absolute value of the calculated order parameter,
    $\abs{\psi}$, shown on a plane sliced perpendicular to $c$-axis
    for nanorods (on the left) and dislocations (on the right) as pinning
    sites. The applied fields are 3 T (top) and 1.5 T (bottom). The scale
    of the images is the same as in fig.\ \ref{simulation}a.
  } 
  \label{vortex-lattice}
\end{figure}

Figure \ref{vortex-lattice} shows examples of simulation results 
at fields $B$ = 3 T and $B$ = 1.5 T with BZO-rods and dislocations as
pinning sites. It can be seen that the strong pinning force of
BZO-nanorods strongly disturbs the vortex lattice while pinning by
dislocations results in a more regular vortex lattice. The modelled 
pinning sites can be seen in these images as larger circles (nanorods) 
and small dots (dislocations). It is easy to determine whether a 
pinning site is occupied when these images are combined with the 
information about the phase of $\psi$ and the pinscape coordinates. 

\begin{figure}
  \includegraphics[width=8.6cm]{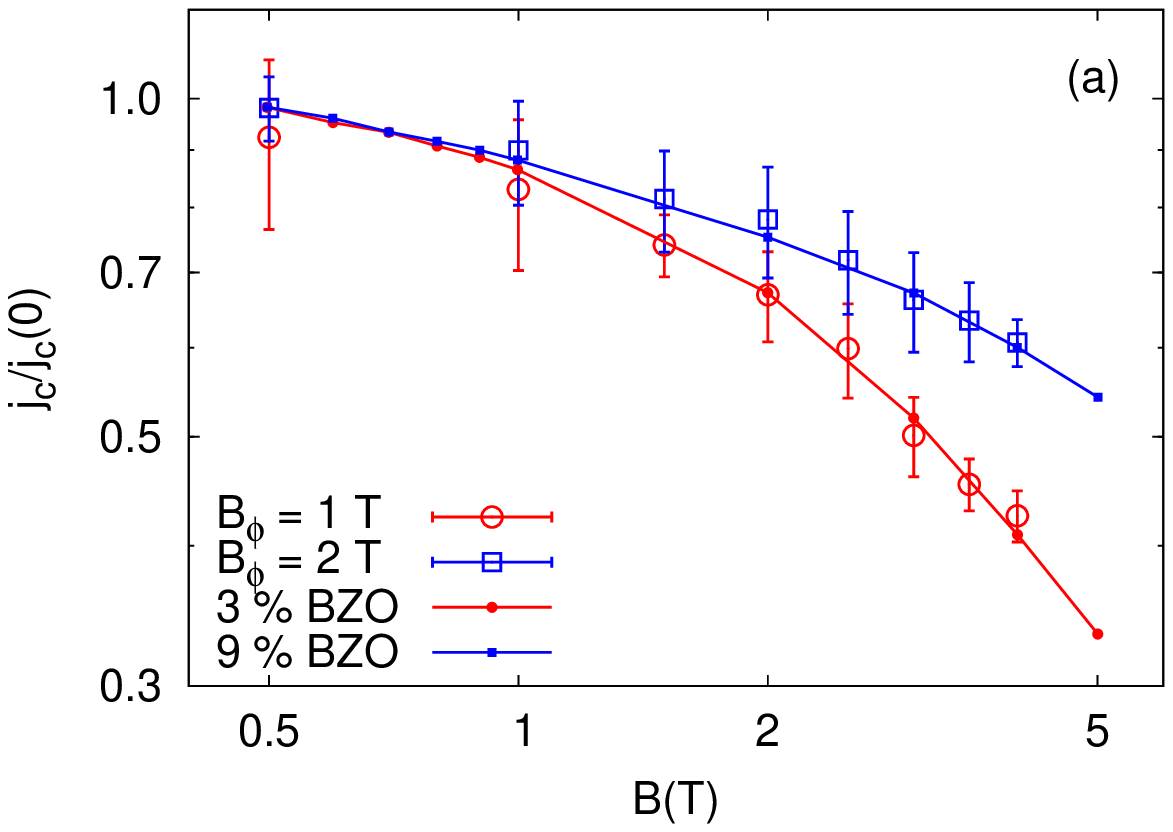}
  \includegraphics[width=8.6cm]{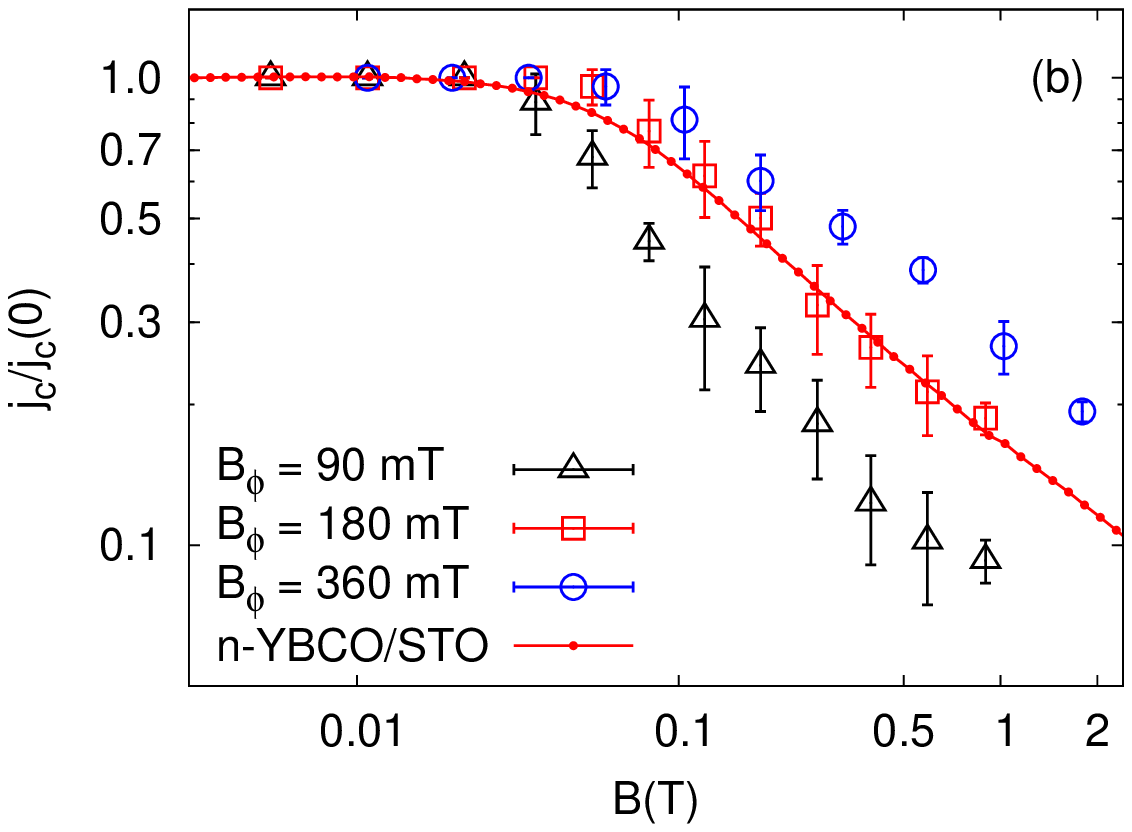}
  \caption{
    Calculated \jc\ (empty symbols) compared to experimental data 
    (linespoints) when the simulation and sample contain a) nanorods
    and b) dislocations. The simulation point is an average over
    several different pinscapes and the errorbars are the standard
    deviations of those.
  } 
  \label{jc-exp}
\end{figure}

The magnetic field scan from high field to low field was calculated
using several random pinscapes with the same number of pinning
sites. The results were taken as the averages of the critical currents 
derived from these simulations and the errors as the standard 
deviations of the simulation sets. In figure \ref{jc-exp} we have 
overlayed the obtained \jc$(B)$ for nanorods with 2.9 wt-\% and 9 wt-\% 
BZO-doped YBCO film data \cite{Peurla3} and the \jc$(B)$ for dislocations 
with data from an undoped YBCO film. The \jc\ values have 
been scaled to 1 at zero field. The agreement with the experimental data is 
excellent and shows that the nanorods are so dominant in pinning that other 
pinning types can be neglected when the rods and the magnetic field are 
parallel. The $\alpha$ values for dislocation simulations range from 0.5 to 
0.8 while for BZO-nanorods simulations gave $\alpha$ values of 0.26 and 
0.41 for rod densities of 1 T and 2 T, respectively, which are quite 
typical values 
\cite{Klaassen2,Peurla3,Peurla1,Huhtinen12,Peurla4,Huhtinen16,Huhtinen20}. 
The $\alpha$ value of the measured pure YBCO 
thin film shown in fig.\ \ref{jc-exp}b is 0.59. Comparing the dislocation 
data to the experiments is more challenging because experimentally 
dislocation density is not so easy to adjust with the growth conditions 
while the density of the nanorods can be adjusted with the doping level. 
Further difficulties arise from the fact that dislocations do not dominate 
pinning so clearly as the large nanorods do. Thus, including other types of 
pinning sites (e.g.\ twins, oxygen vacancies) to the model is needed for 
more in depth analysis of the undoped films.

The nanorod data in fig.\ \ref{jc-exp}a shows a rounded shape of \jc
$(B)$ on log-log scale which is in contrast to the sharp bend between 
the different field regimes of the dislocation data in fig.\ \ref{jc-exp}b. 
Fitting $B^{-\alpha}$ to the rounded curves is difficult because the shape 
of the curve is not really correct unlike with dislocations.
In this work $\alpha$ was determined from the part close to $B^*$ if the 
curve is rounded. 

\begin{figure}
  \includegraphics[width=8.6cm]{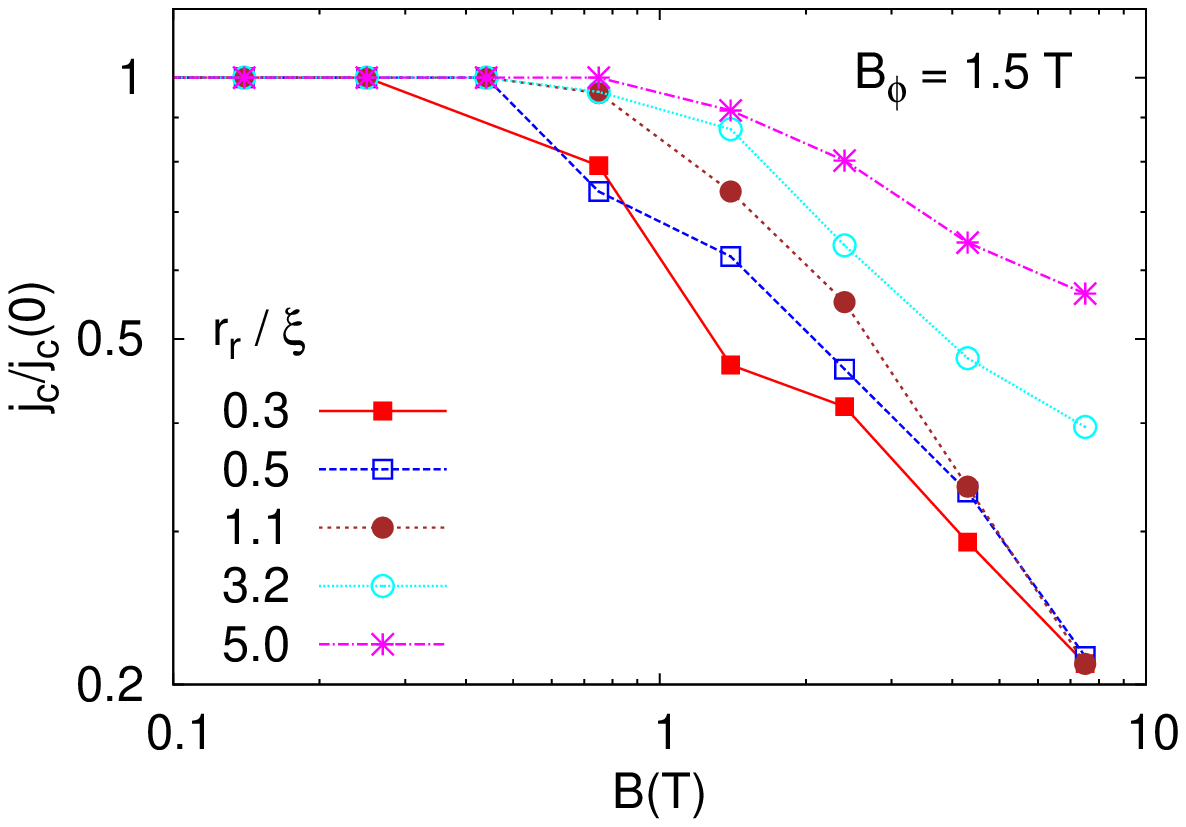}
  \caption{
    Calculated \jc\ with the same rod locations but different
    rod radii $r_\mathrm{r}$.
  }
  \label{jc-vs-r}
\end{figure}

It would seem obvious to attribute the rounded shape of the BZO-nanorod 
data to the more dominant vortex-vortex interactions 
caused by the higher density of the defects. But the dislocations differ 
from nanorods also by size not only by density. Thus, we calculated
\jc\ for a high density ($B_\phi = 1.5$ T) pinscape with fixed defect 
locations but with variable rod radius $r_\mathrm{r}$ which is shown in 
fig.\ \ref{jc-vs-r}. From this it is clear that the typical rounded shape of 
the \jc-curve of the BZO-nanorods requires not only a high density but also 
a large rod size. Having a high density of dislocations will not make the 
sample as good as one with a high density of BZO-nanorods. 

\begin{figure}
  \includegraphics[width=8.6cm]{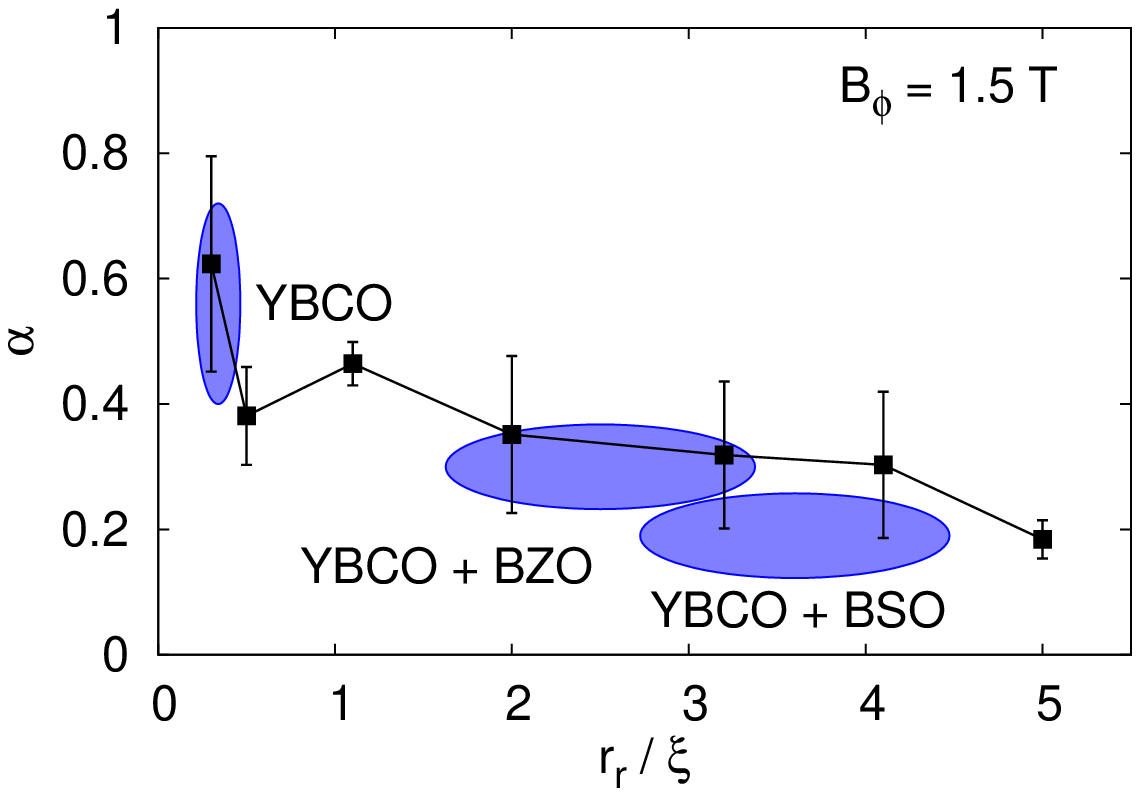}
  \caption{
    The $\alpha$ values from fits to curves in
    fig.\ \ref{jc-vs-r}. The error bars are the standard deviations
    of the fits. The range of experimental values measured for pure
    \cite{Peurla1, Huhtinen12}, BZO doped \cite{Peurla4, Huhtinen16,
    Huhtinen20} and BSO \cite{Varanasi2,Varanasi5} doped YBCO thin films is
    indicated with ellipses. 
  }
  \label{alpha-vs-r}
\end{figure}

The results in fig.\ \ref{jc-vs-r} were further analyzed by fitting \jc
$(B) = cB^{-\alpha}$ to the above $B^*$ portion of the datapoints. The 
obtained $\alpha$ values are shown in fig.\ \ref{alpha-vs-r}. The values 
decrease with increasing rod diameter from 0.6 to 0.2 which is within 
the range of measured values for BZO and BSO nanorods schematically shown 
as ellipses in fig.\ \ref{alpha-vs-r}. Naturally, the decrease of $\alpha$ 
comes from the deeper pinning potential of a larger rod which makes it 
possible for a vortex to sit in the potential well up to higher magnetic 
fields. The ellipses fit the simulation results even better if we take into 
account that in real samples nanorods are surrounded by additional defects 
which makes the rod effectively larger than the actual rod size measured 
with TEM and would thus move the ellipses to the right.

\begin{figure}
  \includegraphics[width=8.6cm]{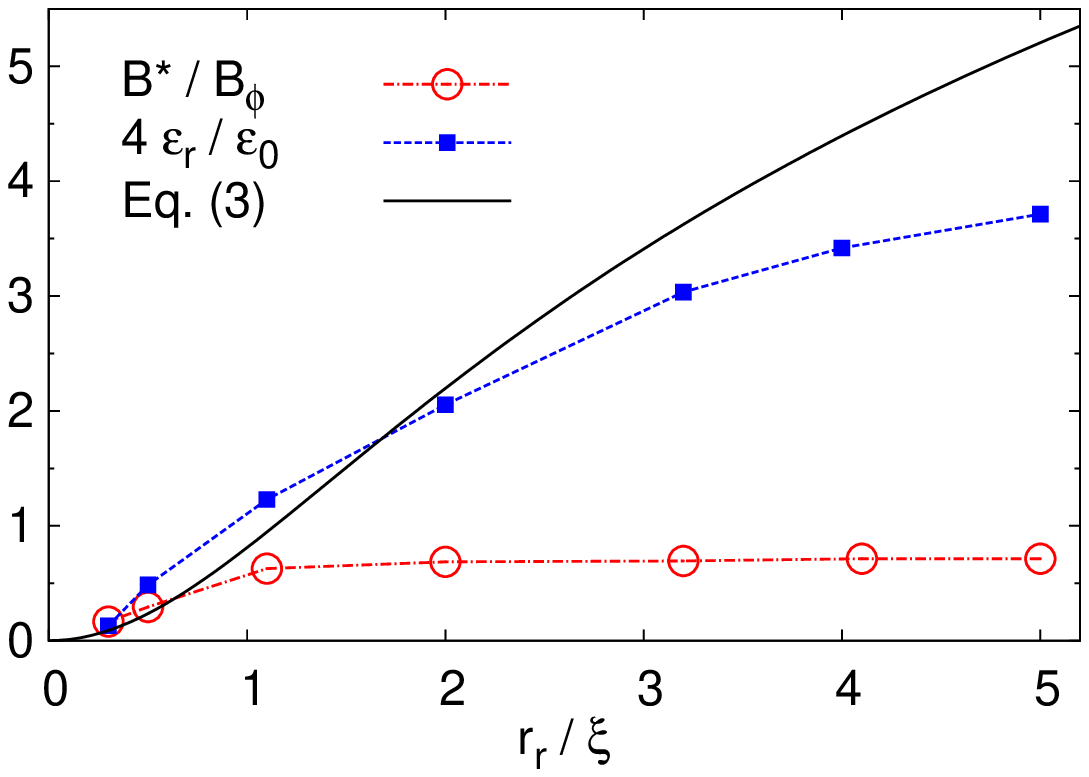}
  \caption{
    The depth of the pinning potential 
    $4\epsilon_\mathrm{r}/\epsilon_0$ as a function of
    the rod radius $r_\mathrm{r}$ as calculated from the
    simulations (blue squares) and as given by eq. \ref{eq:potential}
    (solid curve). The red circles show the ratio of the accommodation
    field $B^*$ to the matching field $B_\phi$ determined from the
    data shown in fig.\ \ref{jc-vs-r}.
  }
  \label{Bstar-vs-r}
\end{figure}

Furthermore, the accommodation field was calculated from the fit as 
$B^*=\mbox{}^\alpha\!\!\sqrt{c}$ i.e.\ the point where  $cB^{-\alpha}$ 
intersects the line \jc$(0)=1$. The accommodation fields relative to the 
matching field are shown in fig.\ \ref{Bstar-vs-r}. 
The accommodation fields level off to value $B^*/B_\phi = 0.7$ starting 
from the rod size $r_\mathrm{r}/\xi=2$ which is the size where several 
vortices get pinned at a single rod. The $B^*/B_\phi = 0.7$ is also seen in 
experiments \cite{Dam2,Klaassen2}.
Assuming strong pinning by linear defects in low magnetic field the 
pinning potential per unit length $\epsilon_\mathrm{r}$ of the vortex is 
related to the accommodation field \cite{Blatter4}:
\begin{equation}
\frac{B^*}{B_\phi} \approx 4\frac{\epsilon_\mathrm{r}}{\epsilon_0},
\label{eq:Bstar}
\end{equation}
where $\epsilon_0 = \pi\hbar^2/(\mu_0q^2\lambda^2)$ is the 
characteristic energy of the vortex per unit length. An analytical 
expression for the depth of the pinning potential of a cylindrical 
cavity has been derived in ref.\ \onlinecite{Blatter4} as an upper limit 
within the London approximation: 
\begin{equation}
4\frac{\epsilon_\mathrm{r}}{\epsilon_0} \approx
2\ln \left( 1+\frac{r_\mathrm{r}^2}{2\xi} \right).
\label{eq:potential}
\end{equation}
Using eqs. \ref{eq:Bstar} and \ref{eq:potential} the accommodation field 
can be roughly related to the pinning potential of the nanorods.
But as can be seen from fig.\ \ref{Bstar-vs-r} this relation between the 
accommodation field and the pinning potential seems to break down at large 
rod sizes where several vortices are pinned per rod.

The pinning potential of a nanorod was also determined by simulating a 
system consisting of a single vortex and a single nanorod. The depth of 
the potential was taken as the difference in the total energy between 
the state where the vortex is far away from the pinning site and where 
the vortex is pinned in the nanorod divided by the length of the rod. 
This was done for several rodsizes and the result is shown in fig.\ \ref
{Bstar-vs-r} where eq. \ref{eq:potential} and the accommodation fields 
are also shown. Considering that there is no scaling or fitting involved 
the calculated potential depths are in general agreement with eq. 
\ref{eq:potential}.

\begin{figure}
  \includegraphics[width=8.6cm]{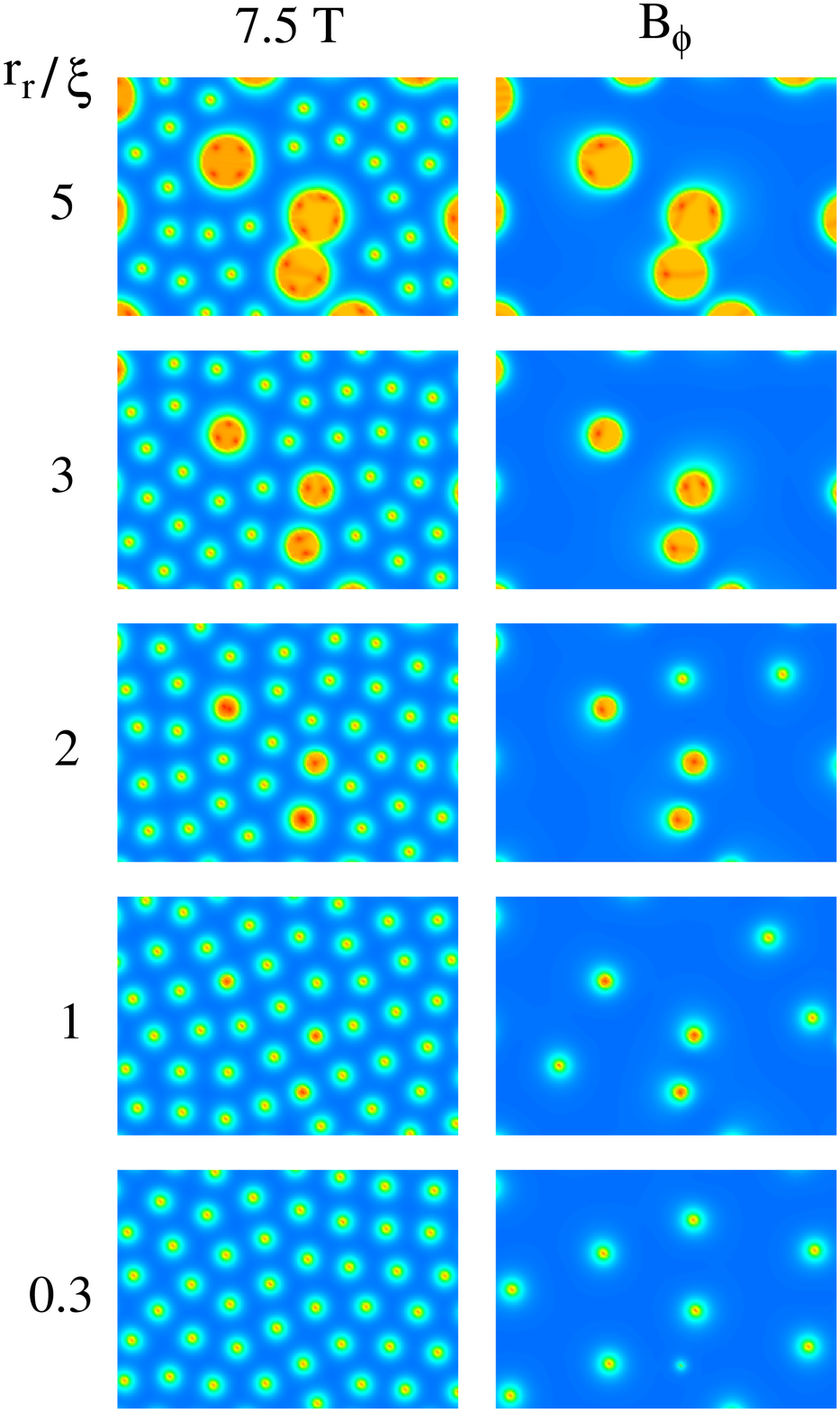}
  \caption{
    A closer look at 3 of the 55 nanorods in the simulations with varying 
    rod size. The absolute value of the order parameter is shown with the 
    color scale from 1 (blue) to 0 (red). The size of the images is 230 $
    \times$ 160 simulation grid points (130 nm $\times$ 90 nm). 
    Simulation results in high magnetic field are on the left and at 
    matching field (1.5 T) are on the right. The size of the nanorod
    decreases from top to bottom.
  }
  \label{rods}
\end{figure}

Examples of more than one vortex pinned to a nanorod are shown in fig.\
\ref{rods}. Up to four vortices get pinned to a large nanorod at high 
magnetic field. The vortices sit symmetrically at opposite sides of the 
rod. At rod size $r_\mathrm{r}/\xi=2$ the vortices are forced to be very 
close to each other but still there are two vortices in two of the rods. At 
rod size $r_\mathrm{r}/\xi=1$ there is only one vortex per rod.
Ideally, there should be one vortex at each pinning site at the matching 
field. But the pinning sites are randomly distributed which makes them 
unevenly spaced. Thus, at small rod sizes the free vortices and the 
unoccupied pinning sites cancel out each other in the large scale giving 
the average of one vortex per pinning site. At large rod sizes this 
cancelling out happens by pinning several vortices at suitable sites while 
leaving some sites empty.

\begin{figure}
  \includegraphics[width=8.6cm]{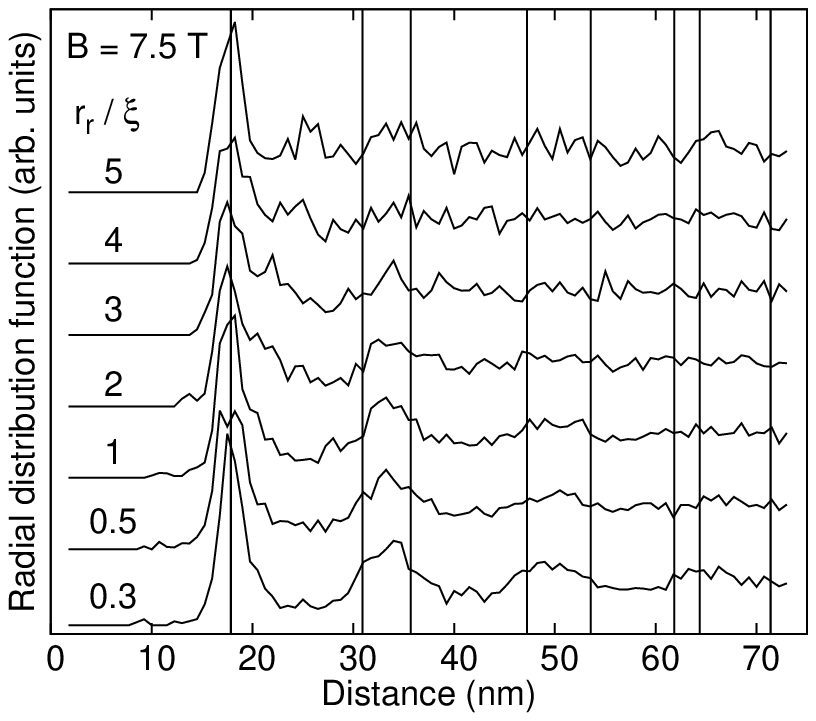}
  \caption{
    The radial distribution functions of the vortex core positions for 
    the different nanorod sizes. The vertical lines mark the positions 
    of the nearest neighbours for an ideal 2D-triangular vortex lattice
    at 7.5 T. 
  }
  \label{rdf}
\end{figure}

It is obvious from fig.\ \ref{rods} that the vortex lattice is more regular 
with the small nanorods. For a closer look into the short and long range 
order the radial distribution functions (fig.\ \ref{rdf}) of the vortex 
positions ($N\approx 250$) were calculated from the simulation results at 
7.5 T. The vertical lines show the positions of the maxima for an ideal
2D-triangular lattice with the unit length of
$a=(2\Phi_0/(\sqrt{3}B))^{1/2}$ where $\Phi_0$ is the flux quantum and
$B=7.5$ T. There are 4 broad peaks visible at the rod size 0.3 which 
roughly overlap the positions of the ideal case. The 4th peak corresponding 
the 6th and 7th nearest neighbour (NN) is already very weak. There is no 
long range order past the 7th NN at any rod size. At the rod size
$r_\mathrm{r}/\xi=3$ and larger only the first NN peak is visible. Thus, 
the triangular vortex lattice continues over the small rods with some 
disturbance while the large rods completely break down the long range 
order. Since the area of the first peak is constant the coordination number 
is the same for all rod sizes. The pinning of several vortices to the same 
pinning site has been experimentally observed e.g.\ in ref.\ 
\onlinecite{Bezryadin2} where the irregularity of the vortex lattice near 
the pinning sites is clearly visible too.
 
From the experimental point of view the results presented here further 
emphasize that one should not only focus on optimizing the density of the 
pinning sites but the size of the defects needs to be carefully considered 
too. Naturally, if the magnetic field in the application is not homogenous 
one needs to consider the splay of the nanorods too. In large scale 
production, where \textit{in situ} deposition is almost the only option,
the size of the defect is fixed by the chemical properties of the dopant
\cite{MacManusDriscoll8}. Thus, one has to try different dopant materials 
to find those that produce large enough columnar defects. With large 
defects the optimal density is a compromise between the superconducting 
volume and pinning. However, at large defect sizes, where several vortices 
get pinned to each defect, it is clear that any density (in terms of 
matching field) of the defects above the operating field of the application 
is not optimal since some of the defects will be left empty. On the other 
hand, if there are two vortices per rod, a density of half the operating 
field is too low because some of the rods will only pin one vortex due to 
e.g.\ the variation in the rod positions. With naive arguments one can say 
that having one vortex per rod is better than two since with two vortices 
the rod diameter has to be roughly twice as large which increases the 
volume of the rods by a factor of 4. But in reality there is a lot of 
different strains and relaxations involved when the distance between the 
rods is changed which could compensate for the loss of the superconducting 
volume by increasing \Tc\ and \jc(0).

\section{Conclusions}

In this paper we have shown that it is possible to model flux pinning 
numerically with the Ginzburg-Landau equations. The model fully includes 
vortex-vortex interactions which is very important for modelling flux 
pinning. Doing this allows us to see the vortex paths inside the sample and 
to calculate the field dependence of \jc\ for a pre-determined pinning site 
configurations. The results are in excellent agreement with the experiments.

The main result was that the reason behind the characteristic round shape 
of the \jc\ and the lowered $\alpha$ value of the BZO-doped YBCO films is 
not only the high density but also the large size of the pinning sites. The 
density of the pinning sites acts through the vortex-vortex interactions 
while the rod size (i) changes the depth of the pinning potential,
(ii) changes the number of the pinned vortices per rod,
(iii) and supresses the long range order of the vortex 
lattice at larger sizes.

The current model can also be used to simulate the angular depedency of
\jc\ which will be our next objective. Adding time and current to the model
by using time-dependent Ginzburg-Landau will allow transport measurement
simulations.

\begin{acknowledgments}
  The authors thank CSC~--~Scientific Computing Ltd. for generous
  supercomputer resources. The Wihuri Foundation is acknowledged for 
  financial support. JJ also acknowledges UK EPSRC grant for
  support.
\end{acknowledgments}

\bibliographystyle{apsrev4-1}

\begin{thebibliography}{10}%
\makeatletter
\providecommand \@ifxundefined [1]{%
 \ifx #1\undefined \expandafter \@firstoftwo
 \else \expandafter \@secondoftwo
\fi
}%
\providecommand \@ifnum [1]{%
 \ifnum #1\expandafter \@firstoftwo
 \else \expandafter \@secondoftwo
\fi
}%
\providecommand \enquote [1]{``#1''}%
\providecommand \bibnamefont  [1]{#1}%
\providecommand \bibfnamefont [1]{#1}%
\providecommand \citenamefont [1]{#1}%
\providecommand\href[0]{\@sanitize\@href}%
\providecommand\@href[1]{\endgroup\@@startlink{#1}\endgroup\@@href}%
\providecommand\@@href[1]{#1\@@endlink}%
\providecommand \@sanitize [0]{\begingroup\catcode`\&12\catcode`\#12\relax}%
\@ifxundefined \pdfoutput {\@firstoftwo}{%
 \@ifnum{\z@=\pdfoutput}{\@firstoftwo}{\@secondoftwo}%
}{%
 \providecommand\@@startlink[1]{\leavevmode}%
 \providecommand\@@endlink[0]{}%
}{%
 \providecommand\@@startlink[1]{%
  \leavevmode
  \pdfstartlink
   attr{/Border[0 0 1 ]/H/I/C[0 1 1]}%
   user{/Subtype/Link/A<</Type/Action/S/URI/URI(#1)>>}%
  \relax
 }%
 \providecommand\@@endlink[0]{\pdfendlink}%
}%
\providecommand \url  [0]{\begingroup\@sanitize \@url }%
\providecommand \@url [1]{\endgroup\@href {#1}{\urlprefix}}%
\providecommand \urlprefix [0]{URL }%
\providecommand \Eprint[0]{\href }%
\@ifxundefined \urlstyle {%
  \providecommand \doi [1]{doi:\discretionary{}{}{}#1}%
}{%
  \providecommand \doi [0]{doi:\discretionary{}{}{}\begingroup
  \urlstyle{rm}\Url }%
}%
\providecommand \doibase [0]{http://dx.doi.org/}%
\providecommand \Doi[1]{\href{\doibase#1}}%
\providecommand \bibAnnote [3]{%
  \BibitemShut{#1}%
  \begin{quotation}\noindent
    \textsc{Key:}\ #2\\\textsc{Annotation:}\ #3%
  \end{quotation}%
}%
\providecommand \bibAnnoteFile [2]{%
  \IfFileExists{#2}{\bibAnnote {#1} {#2} {\input{#2}}}{}%
}%
\providecommand \typeout [0]{\immediate \write \m@ne }%
\providecommand \selectlanguage [0]{\@gobble}%
\providecommand \bibinfo [0]{\@secondoftwo}%
\providecommand \bibfield [0]{\@secondoftwo}%
\providecommand \translation [1]{[#1]}%
\providecommand \BibitemOpen[0]{}%
\providecommand \bibitemStop [0]{}%
\providecommand \bibitemNoStop [0]{.\EOS\space}%
\providecommand \EOS [0]{\spacefactor3000\relax}%
\providecommand \BibitemShut [1]{\csname bibitem#1\endcsname}%
\bibitem{Foltyn4}%
  \BibitemOpen
  \bibfield{author}{%
  \bibinfo {author} {\bibfnamefont{S.~R.}\ \bibnamefont{Foltyn}}, \bibinfo
  {author} {\bibfnamefont{L.}~\bibnamefont{Civale}}, \bibinfo {author}
  {\bibfnamefont{J.~L.}\ \bibnamefont{MacManus-Driscoll}}, \bibinfo {author}
  {\bibfnamefont{Q.~X.}\ \bibnamefont{Jia}}, \bibinfo {author}
  {\bibfnamefont{B.}~\bibnamefont{Maiorov}}, \bibinfo {author}
  {\bibfnamefont{H.}~\bibnamefont{Wang}},\ and\ \bibinfo {author}
  {\bibfnamefont{M.}~\bibnamefont{Maley}},\ }%
  \bibfield{journal}{%
  \bibinfo {journal} {Nat. Mater.}\ }%
  \textbf{\bibinfo {volume} {6}},\ \bibinfo {pages} {631} (\bibinfo {year}
  {2007})%
  \bibAnnoteFile{NoStop}{Foltyn4}%
\bibitem{Pan3}%
  \BibitemOpen
  \bibfield{author}{%
  \bibinfo {author} {\bibfnamefont{V.}~\bibnamefont{Pan}}, \bibinfo {author}
  {\bibfnamefont{Y.}~\bibnamefont{Cherpak}}, \bibinfo {author}
  {\bibfnamefont{V.}~\bibnamefont{Komashko}}, \bibinfo {author}
  {\bibfnamefont{S.}~\bibnamefont{Pozigun}}, \bibinfo {author}
  {\bibfnamefont{C.}~\bibnamefont{Tretiatchenko}}, \bibinfo {author}
  {\bibfnamefont{A.}~\bibnamefont{Semenov}}, \bibinfo {author}
  {\bibfnamefont{E.}~\bibnamefont{Pashitskii}},\ and\ \bibinfo {author}
  {\bibfnamefont{A.~V.}\ \bibnamefont{Pan}},\ }%
  \bibfield{journal}{%
  \bibinfo {journal} {Phys. Rev. B}\ }%
  \textbf{\bibinfo {volume} {73}},\ \bibinfo {pages} {054508} (\bibinfo {year}
  {2006})%
  \bibAnnoteFile{NoStop}{Pan3}%
\bibitem{Dam2}%
  \BibitemOpen
  \bibfield{author}{%
  \bibinfo {author} {\bibfnamefont{B.}~\bibnamefont{Dam}}, \bibinfo {author}
  {\bibfnamefont{J.~M.}\ \bibnamefont{Huijbregtse}}, \bibinfo {author}
  {\bibfnamefont{F.~C.}\ \bibnamefont{Klaassen}}, \bibinfo {author}
  {\bibfnamefont{R.~C.~F.}\ \bibnamefont{van~der Geest}}, \bibinfo {author}
  {\bibfnamefont{G.}~\bibnamefont{Doornbos}}, \bibinfo {author}
  {\bibfnamefont{J.~H.}\ \bibnamefont{Rector}}, \bibinfo {author}
  {\bibfnamefont{A.~M.}\ \bibnamefont{Testa}}, \bibinfo {author}
  {\bibfnamefont{S.}~\bibnamefont{Freisem}}, \bibinfo {author}
  {\bibfnamefont{J.~C.}\ \bibnamefont{Martinez}}, \bibinfo {author}
  {\bibfnamefont{B.}~\bibnamefont{Stuble-Pumpin}},\ and\ \bibinfo {author}
  {\bibfnamefont{R.}~\bibnamefont{Griessen}},\ }%
  \bibfield{journal}{%
  \bibinfo {journal} {Nature}\ }%
  \textbf{\bibinfo {volume} {399}},\ \bibinfo {pages} {439} (\bibinfo {year}
  {1999})%
  \bibAnnoteFile{NoStop}{Dam2}%
\bibitem{Svetchnikov1}%
  \BibitemOpen
  \bibfield{author}{%
  \bibinfo {author} {\bibfnamefont{V.}~\bibnamefont{Svetchnikov}}, \bibinfo
  {author} {\bibfnamefont{V.}~\bibnamefont{Pan}}, \bibinfo {author}
  {\bibfnamefont{C.}~\bibnamefont{Tr{\ae}holt}},\ and\ \bibinfo {author}
  {\bibfnamefont{H.}~\bibnamefont{Zandbergen}},\ }%
  \bibfield{journal}{%
  \bibinfo {journal} {IEEE T. Appl. Supercond.}\ }%
  \textbf{\bibinfo {volume} {7}},\ \bibinfo {pages} {1396} (\bibinfo {year}
  {1997})%
  \bibAnnoteFile{NoStop}{Svetchnikov1}%
\bibitem{Blatter4}%
  \BibitemOpen
  \bibfield{author}{%
  \bibinfo {author} {\bibfnamefont{G.}~\bibnamefont{Blatter}}, \bibinfo
  {author} {\bibfnamefont{M.}~\bibnamefont{Feigel'man}}, \bibinfo {author}
  {\bibfnamefont{V.}~\bibnamefont{Geshkenbein}}, \bibinfo {author}
  {\bibfnamefont{A.}~\bibnamefont{Larkin}},\ and\ \bibinfo {author}
  {\bibfnamefont{V.}~\bibnamefont{Vinokur}},\ }%
  \bibfield{journal}{%
  \bibinfo {journal} {Rev. Mod. Phys.}\ }%
  \textbf{\bibinfo {volume} {66}},\ \bibinfo {pages} {1125} (\bibinfo {year}
  {1994})%
  \bibAnnoteFile{NoStop}{Blatter4}%
\bibitem{Crabtree1}%
  \BibitemOpen
  \bibfield{author}{%
  \bibinfo {author} {\bibfnamefont{G.~W.}\ \bibnamefont{Crabtree}}, \bibinfo
  {author} {\bibfnamefont{D.~O.}\ \bibnamefont{Gunter}}, \bibinfo {author}
  {\bibfnamefont{H.~G.}\ \bibnamefont{Kaper}}, \bibinfo {author}
  {\bibfnamefont{A.~E.}\ \bibnamefont{Koshelev}}, \bibinfo {author}
  {\bibfnamefont{G.~K.}\ \bibnamefont{Leaf}},\ and\ \bibinfo {author}
  {\bibfnamefont{V.~M.}\ \bibnamefont{Vinokur}},\ }%
  \bibfield{journal}{%
  \bibinfo {journal} {Phys. Rev. B}\ }%
  \textbf{\bibinfo {volume} {61}},\ \bibinfo {pages} {1446} (\bibinfo {year}
  {2000})%
  \bibAnnoteFile{NoStop}{Crabtree1}%
\bibitem{MacManusDriscoll2}%
  \BibitemOpen
  \bibfield{author}{%
  \bibinfo {author} {\bibfnamefont{J.~L.}\ \bibnamefont{MacManus-Driscoll}},
  \bibinfo {author} {\bibfnamefont{S.~R.}\ \bibnamefont{Foltyn}}, \bibinfo
  {author} {\bibfnamefont{Q.~X.}\ \bibnamefont{Jia}}, \bibinfo {author}
  {\bibfnamefont{H.}~\bibnamefont{Wang}}, \bibinfo {author}
  {\bibfnamefont{A.}~\bibnamefont{Serquis}}, \bibinfo {author}
  {\bibfnamefont{L.}~\bibnamefont{Civale}}, \bibinfo {author}
  {\bibfnamefont{B.}~\bibnamefont{Maiorov}}, \bibinfo {author}
  {\bibfnamefont{M.~E.}\ \bibnamefont{Hawley}}, \bibinfo {author}
  {\bibfnamefont{M.~P.}\ \bibnamefont{Maley}},\ and\ \bibinfo {author}
  {\bibfnamefont{D.~E.}\ \bibnamefont{Peterson}},\ }%
  \bibfield{journal}{%
  \bibinfo {journal} {Nat. Mater.}\ }%
  \textbf{\bibinfo {volume} {3}},\ \bibinfo {pages} {439} (\bibinfo {year}
  {2004})%
  \bibAnnoteFile{NoStop}{MacManusDriscoll2}%
\bibitem{MacManusDriscoll8}%
  \BibitemOpen
  \bibfield{author}{%
  \bibinfo {author} {\bibfnamefont{J.~L.}\ \bibnamefont{MacManus-Driscoll}},
  \bibinfo {author} {\bibfnamefont{S.~A.}\ \bibnamefont{Harrington}}, \bibinfo
  {author} {\bibfnamefont{J.~H.}\ \bibnamefont{Durrell}}, \bibinfo {author}
  {\bibfnamefont{G.}~\bibnamefont{Ercolano}}, \bibinfo {author}
  {\bibfnamefont{H.}~\bibnamefont{Wang}}, \bibinfo {author}
  {\bibfnamefont{J.~H.}\ \bibnamefont{Lee}}, \bibinfo {author}
  {\bibfnamefont{C.~F.}\ \bibnamefont{Tsai}}, \bibinfo {author}
  {\bibfnamefont{B.}~\bibnamefont{Maiorov}}, \bibinfo {author}
  {\bibfnamefont{A.}~\bibnamefont{Kursumovic}},\ and\ \bibinfo {author}
  {\bibfnamefont{S.~C.}\ \bibnamefont{Wimbush}},\ }%
  \bibfield{journal}{%
  \bibinfo {journal} {Supercond. Sci. Technol.}\ }%
  \textbf{\bibinfo {volume} {23}},\ \bibinfo {pages} {034009} (\bibinfo {year}
  {2010})%
  \bibAnnoteFile{NoStop}{MacManusDriscoll8}%
\bibitem{Varanasi2}%
  \BibitemOpen
  \bibfield{author}{%
  \bibinfo {author} {\bibfnamefont{C.~V.}\ \bibnamefont{Varanasi}}, \bibinfo
  {author} {\bibfnamefont{P.~N.}\ \bibnamefont{Barnes}}, \bibinfo {author}
  {\bibfnamefont{J.}~\bibnamefont{Burke}}, \bibinfo {author}
  {\bibfnamefont{L.}~\bibnamefont{Brunke}}, \bibinfo {author}
  {\bibfnamefont{I.}~\bibnamefont{Maartense}}, \bibinfo {author}
  {\bibfnamefont{T.~J.}\ \bibnamefont{Haugan}}, \bibinfo {author}
  {\bibfnamefont{E.~A.}\ \bibnamefont{Stinzianni}}, \bibinfo {author}
  {\bibfnamefont{K.~A.}\ \bibnamefont{Dunn}},\ and\ \bibinfo {author}
  {\bibfnamefont{P.}~\bibnamefont{Haldar}},\ }%
  \bibfield{journal}{%
  \bibinfo {journal} {Supercond. Sci. Technol.}\ }%
  \textbf{\bibinfo {volume} {19}},\ \bibinfo {pages} {L37} (\bibinfo {year}
  {2006})%
  \bibAnnoteFile{NoStop}{Varanasi2}%
\bibitem{Varanasi5}%
  \BibitemOpen
  \bibfield{author}{%
  \bibinfo {author} {\bibfnamefont{C.~V.}\ \bibnamefont{Varanasi}}, \bibinfo
  {author} {\bibfnamefont{J.}~\bibnamefont{Burke}}, \bibinfo {author}
  {\bibfnamefont{L.}~\bibnamefont{Brunke}}, \bibinfo {author}
  {\bibfnamefont{H.}~\bibnamefont{Wang}}, \bibinfo {author}
  {\bibfnamefont{M.}~\bibnamefont{Sumption}},\ and\ \bibinfo {author}
  {\bibfnamefont{P.~N.}\ \bibnamefont{Barnes}},\ }%
  \bibfield{journal}{%
  \bibinfo {journal} {J. Appl. Phys.}\ }%
  \textbf{\bibinfo {volume} {102}},\ \bibinfo {pages} {063909} (\bibinfo {year}
  {2007})%
  \bibAnnoteFile{NoStop}{Varanasi5}%
\bibitem{Peurla3}%
  \BibitemOpen
  \bibfield{author}{%
  \bibinfo {author} {\bibfnamefont{M.}~\bibnamefont{Peurla}}, \bibinfo {author}
  {\bibfnamefont{P.}~\bibnamefont{Paturi}}, \bibinfo {author}
  {\bibfnamefont{Y.~P.}\ \bibnamefont{Stepanov}}, \bibinfo {author}
  {\bibfnamefont{H.}~\bibnamefont{Huhtinen}}, \bibinfo {author}
  {\bibfnamefont{Y.~Y.}\ \bibnamefont{Tse}}, \bibinfo {author}
  {\bibfnamefont{A.~C.}\ \bibnamefont{B{\'o}di}}, \bibinfo {author}
  {\bibfnamefont{J.}~\bibnamefont{Raittila}},\ and\ \bibinfo {author}
  {\bibfnamefont{R.}~\bibnamefont{Laiho}},\ }%
  \bibfield{journal}{%
  \bibinfo {journal} {Supercond. Sci. Technol.}\ }%
  \textbf{\bibinfo {volume} {19}},\ \bibinfo {pages} {767} (\bibinfo {year}
  {2006})%
  \bibAnnoteFile{NoStop}{Peurla3}%
\bibitem{Augieri5}%
  \BibitemOpen
  \bibfield{author}{%
  \bibinfo {author} {\bibfnamefont{A.}~\bibnamefont{Augieri}}, \bibinfo
  {author} {\bibfnamefont{G.}~\bibnamefont{Celentano}}, \bibinfo {author}
  {\bibfnamefont{V.}~\bibnamefont{Galluzzi}}, \bibinfo {author}
  {\bibfnamefont{A.}~\bibnamefont{Mancini}}, \bibinfo {author}
  {\bibfnamefont{A.}~\bibnamefont{Rufoloni}}, \bibinfo {author}
  {\bibfnamefont{A.}~\bibnamefont{Vannozzi}}, \bibinfo {author}
  {\bibfnamefont{A.~A.}\ \bibnamefont{Armenio}}, \bibinfo {author}
  {\bibfnamefont{T.}~\bibnamefont{Petrisor}}, \bibinfo {author}
  {\bibfnamefont{L.}~\bibnamefont{Ciontea}}, \bibinfo {author}
  {\bibfnamefont{S.}~\bibnamefont{Rubanov}}, \bibinfo {author}
  {\bibfnamefont{E.}~\bibnamefont{Silva}},\ and\ \bibinfo {author}
  {\bibfnamefont{N.}~\bibnamefont{Pompeo}},\ }%
  \bibfield{journal}{%
  \bibinfo {journal} {J. Appl. Phys.}\ }%
  \textbf{\bibinfo {volume} {108}},\ \bibinfo {pages} {063906} (\bibinfo {year}
  {2010})%
  \bibAnnoteFile{NoStop}{Augieri5}%
\bibitem{Paturi15}%
  \BibitemOpen
  \bibfield{author}{%
  \bibinfo {author} {\bibfnamefont{P.}~\bibnamefont{Paturi}}, \bibinfo {author}
  {\bibfnamefont{M.}~\bibnamefont{Irjala}},\ and\ \bibinfo {author}
  {\bibfnamefont{H.}~\bibnamefont{Huhtinen}},\ }%
  \bibfield{journal}{%
  \bibinfo {journal} {J. Appl. Phys.}\ }%
  \textbf{\bibinfo {volume} {103}},\ \bibinfo {pages} {123907} (\bibinfo {year}
  {2008})%
  \bibAnnoteFile{NoStop}{Paturi15}%
\bibitem{Goyal4}%
  \BibitemOpen
  \bibfield{author}{%
  \bibinfo {author} {\bibfnamefont{A.}~\bibnamefont{Goyal}}, \bibinfo {author}
  {\bibfnamefont{S.}~\bibnamefont{Kang}}, \bibinfo {author}
  {\bibfnamefont{K.~J.}\ \bibnamefont{Leonard}}, \bibinfo {author}
  {\bibfnamefont{P.~M.}\ \bibnamefont{Martin}}, \bibinfo {author}
  {\bibfnamefont{A.~A.}\ \bibnamefont{Gapud}}, \bibinfo {author}
  {\bibfnamefont{M.}~\bibnamefont{Varela}}, \bibinfo {author}
  {\bibfnamefont{M.}~\bibnamefont{Paranthaman}}, \bibinfo {author}
  {\bibfnamefont{A.~O.}\ \bibnamefont{Ijadoula}}, \bibinfo {author}
  {\bibfnamefont{E.~D.}\ \bibnamefont{Specht}}, \bibinfo {author}
  {\bibfnamefont{J.~R.}\ \bibnamefont{Thompson}}, \bibinfo {author}
  {\bibfnamefont{D.~K.}\ \bibnamefont{Christen}}, \bibinfo {author}
  {\bibfnamefont{S.~J.}\ \bibnamefont{Pennycook}},\ and\ \bibinfo {author}
  {\bibfnamefont{F.~A.}\ \bibnamefont{List}},\ }%
  \bibfield{journal}{%
  \bibinfo {journal} {Supercond. Sci. Technol.}\ }%
  \textbf{\bibinfo {volume} {18}},\ \bibinfo {pages} {1533} (\bibinfo {year}
  {2005})%
  \bibAnnoteFile{NoStop}{Goyal4}%
\bibitem{Paturi17}%
  \BibitemOpen
  \bibfield{author}{%
  \bibinfo {author} {\bibfnamefont{P.}~\bibnamefont{Paturi}}, \bibinfo {author}
  {\bibfnamefont{M.}~\bibnamefont{Irjala}}, \bibinfo {author}
  {\bibfnamefont{A.~B.}\ \bibnamefont{Abrahamsen}},\ and\ \bibinfo {author}
  {\bibfnamefont{H.}~\bibnamefont{Huhtinen}},\ }%
  \bibfield{journal}{%
  \bibinfo {journal} {IEEE T. Appl. Supercond.}\ }%
  \textbf{\bibinfo {volume} {19}},\ \bibinfo {pages} {3431} (\bibinfo {year}
  {2009})%
  \bibAnnoteFile{NoStop}{Paturi17}%
\bibitem{Klaassen2}%
  \BibitemOpen
  \bibfield{author}{%
  \bibinfo {author} {\bibfnamefont{F.~C.}\ \bibnamefont{Klaassen}}, \bibinfo
  {author} {\bibfnamefont{G.}~\bibnamefont{Doornbos}}, \bibinfo {author}
  {\bibfnamefont{J.~M.}\ \bibnamefont{Huijbregtse}}, \bibinfo {author}
  {\bibfnamefont{R.~C.~F.}\ \bibnamefont{van~der Geest}}, \bibinfo {author}
  {\bibfnamefont{B.}~\bibnamefont{Dam}},\ and\ \bibinfo {author}
  {\bibfnamefont{R.}~\bibnamefont{Griessen}},\ }%
  \bibfield{journal}{%
  \bibinfo {journal} {Phys. Rev. B}\ }%
  \textbf{\bibinfo {volume} {64}},\ \bibinfo {pages} {184523} (\bibinfo {year}
  {2001})%
  \bibAnnoteFile{NoStop}{Klaassen2}%
\bibitem{Beek1}%
  \BibitemOpen
  \bibfield{author}{%
  \bibinfo {author} {\bibfnamefont{C.~J.}\ \bibnamefont{van~der Beek}},
  \bibinfo {author} {\bibfnamefont{M.}~\bibnamefont{Konczykowski}}, \bibinfo
  {author} {\bibfnamefont{A.}~\bibnamefont{Abal'oshev}}, \bibinfo {author}
  {\bibfnamefont{I.}~\bibnamefont{Abal'osheva}}, \bibinfo {author}
  {\bibfnamefont{P.}~\bibnamefont{Gierlowski}}, \bibinfo {author}
  {\bibfnamefont{S.~J.}\ \bibnamefont{Lewandowski}}, \bibinfo {author}
  {\bibfnamefont{M.~V.}\ \bibnamefont{Indenbom}},\ and\ \bibinfo {author}
  {\bibfnamefont{S.}~\bibnamefont{Barbanera}},\ }%
  \bibfield{journal}{%
  \bibinfo {journal} {Phys. Rev. B}\ }%
  \textbf{\bibinfo {volume} {66}},\ \bibinfo {pages} {24523} (\bibinfo {year}
  {2002})%
  \bibAnnoteFile{NoStop}{Beek1}%
\bibitem{Nelson4}%
  \BibitemOpen
  \bibfield{author}{%
  \bibinfo {author} {\bibfnamefont{D.~R.}\ \bibnamefont{Nelson}}\ and\ \bibinfo
  {author} {\bibfnamefont{V.~M.}\ \bibnamefont{Vinokur}},\ }%
  \bibfield{journal}{%
  \bibinfo {journal} {Phys. Rev. B}\ }%
  \textbf{\bibinfo {volume} {48}},\ \bibinfo {pages} {13060} (\bibinfo {year}
  {1993})%
  \bibAnnoteFile{NoStop}{Nelson4}%
\bibitem{Paturi16}%
  \BibitemOpen
  \bibfield{author}{%
  \bibinfo {author} {\bibfnamefont{P.}~\bibnamefont{Paturi}}, \bibinfo {author}
  {\bibfnamefont{M.}~\bibnamefont{Irjala}}, \bibinfo {author}
  {\bibfnamefont{H.}~\bibnamefont{Huhtinen}},\ and\ \bibinfo {author}
  {\bibfnamefont{A.~B.}\ \bibnamefont{Abrahamsen}},\ }%
  \bibfield{journal}{%
  \bibinfo {journal} {J. Appl. Phys.}\ }%
  \textbf{\bibinfo {volume} {105}},\ \bibinfo {pages} {023904} (\bibinfo {year}
  {2009})%
  \bibAnnoteFile{NoStop}{Paturi16}%
\bibitem{Paturi18}%
  \BibitemOpen
  \bibfield{author}{%
  \bibinfo {author} {\bibfnamefont{P.}~\bibnamefont{Paturi}},\ }%
  \bibfield{journal}{%
  \bibinfo {journal} {Supercond. Sci. Technol.}\ }%
  \textbf{\bibinfo {volume} {23}},\ \bibinfo {pages} {025030} (\bibinfo {year}
  {2010})%
  \bibAnnoteFile{NoStop}{Paturi18}%
\bibitem{Long4}%
  \BibitemOpen
  \bibfield{author}{%
  \bibinfo {author} {\bibfnamefont{N.~J.}\ \bibnamefont{Long}}, \bibinfo
  {author} {\bibfnamefont{N.~M.}\ \bibnamefont{Strickland}},\ and\ \bibinfo
  {author} {\bibfnamefont{E.~F.}\ \bibnamefont{Talantsev}},\ }%
  \bibfield{journal}{%
  \bibinfo {journal} {IEEE T. Appl. Supercond.}\ }%
  \textbf{\bibinfo {volume} {17}},\ \bibinfo {pages} {3684} (\bibinfo {year}
  {2007})%
  \bibAnnoteFile{NoStop}{Long4}%
\bibitem{Long5}%
  \BibitemOpen
  \bibfield{author}{%
  \bibinfo {author} {\bibfnamefont{N.~J.}\ \bibnamefont{Long}},\ }%
  \bibfield{journal}{%
  \bibinfo {journal} {Supercond. Sci. Technol.}\ }%
  \textbf{\bibinfo {volume} {21}},\ \bibinfo {pages} {025007} (\bibinfo {year}
  {2008})%
  \bibAnnoteFile{NoStop}{Long5}%
\bibitem{BarbaOrtega1}%
  \BibitemOpen
  \bibfield{author}{%
  \bibinfo {author} {\bibfnamefont{J.}~\bibnamefont{Barba-Ortega}}, \bibinfo
  {author} {\bibfnamefont{E.}~\bibnamefont{Sardella}},\ and\ \bibinfo {author}
  {\bibfnamefont{J.~A.}\ \bibnamefont{Aguiar}},\ }%
  \bibfield{journal}{%
  \bibinfo {journal} {Supercond. Sci. Technol.}\ }%
  \textbf{\bibinfo {volume} {24}},\ \bibinfo {pages} {015001} (\bibinfo {year}
  {2011})%
  \bibAnnoteFile{NoStop}{BarbaOrtega1}%
\bibitem{Machida1}%
  \BibitemOpen
  \bibfield{author}{%
  \bibinfo {author} {\bibfnamefont{M.}~\bibnamefont{Machida}}\ and\ \bibinfo
  {author} {\bibfnamefont{H.}~\bibnamefont{Kaburaki}},\ }%
  \bibfield{journal}{%
  \bibinfo {journal} {Phys. Rev. Lett.}\ }%
  \textbf{\bibinfo {volume} {75}},\ \bibinfo {pages} {3178} (\bibinfo {year}
  {1995})%
  \bibAnnoteFile{NoStop}{Machida1}%
\bibitem{Nakai1}%
  \BibitemOpen
  \bibfield{author}{%
  \bibinfo {author} {\bibfnamefont{N.}~\bibnamefont{Nakai}}, \bibinfo {author}
  {\bibfnamefont{N.}~\bibnamefont{Hayashi}},\ and\ \bibinfo {author}
  {\bibfnamefont{M.}~\bibnamefont{Machida}},\ }%
  \bibfield{journal}{%
  \bibinfo {journal} {J. Phys. Chem. Solids}\ }%
  \textbf{\bibinfo {volume} {69}},\ \bibinfo {pages} {3301} (\bibinfo {year}
  {2008})%
  \bibAnnoteFile{NoStop}{Nakai1}%
\bibitem{Jaykka2}%
  \BibitemOpen
  \bibfield{author}{%
  \bibinfo {author} {\bibfnamefont{J.}~\bibnamefont{J{\"a}ykk{\"a}}},\ }%
  \bibfield{journal}{%
  \bibinfo {journal} {Phys. Rev.}\ }%
  \textbf{\bibinfo {volume} {D79}},\ \bibinfo {pages} {065006} (\bibinfo {year}
  {2009})%
  \bibAnnoteFile{NoStop}{Jaykka2}%
\bibitem{tao-user-ref}%
  \BibitemOpen
  \bibfield{author}{%
  \bibinfo {author} {\bibfnamefont{S.}~\bibnamefont{Benson}}, \bibinfo {author}
  {\bibfnamefont{L.~C.}\ \bibnamefont{McInnes}}, \bibinfo {author}
  {\bibfnamefont{J.}~\bibnamefont{Mor\'{e}}}, \bibinfo {author}
  {\bibfnamefont{T.}~\bibnamefont{Munson}},\ and\ \bibinfo {author}
  {\bibfnamefont{J.}~\bibnamefont{Sarich}},\ }%
  \emph{\bibinfo {title} {{TAO} User Manual (Revision 1.9)}},\ \bibinfo {type}
  {Tech. Rep.}\ \bibinfo {number} {ANL/MCS-TM-242}\ (\bibinfo {institution}
  {Mathematics and Computer Science Division, Argonne National Laboratory},\
  \bibinfo {year} {2007})\ \bibinfo {note} {http://www.mcs.anl.gov/tao}%
  \bibAnnoteFile{NoStop}{tao-user-ref}%
\bibitem{petsc-web-page}%
  \BibitemOpen
  \bibfield{author}{%
  \bibinfo {author} {\bibfnamefont{S.}~\bibnamefont{Balay}}, \bibinfo {author}
  {\bibfnamefont{K.}~\bibnamefont{Buschelman}}, \bibinfo {author}
  {\bibfnamefont{W.~D.}\ \bibnamefont{Gropp}}, \bibinfo {author}
  {\bibfnamefont{D.}~\bibnamefont{Kaushik}}, \bibinfo {author}
  {\bibfnamefont{M.~G.}\ \bibnamefont{Knepley}}, \bibinfo {author}
  {\bibfnamefont{L.~C.}\ \bibnamefont{McInnes}}, \bibinfo {author}
  {\bibfnamefont{B.~F.}\ \bibnamefont{Smith}},\ and\ \bibinfo {author}
  {\bibfnamefont{H.}~\bibnamefont{Zhang}},\ }%
  \enquote{\bibinfo {title} {{PETSc} {W}eb page},}\  (\bibinfo {year} {2009}),\
  \bibinfo {note} {http://www.mcs.anl.gov/petsc}%
  \bibAnnoteFile{NoStop}{petsc-web-page}%
\bibitem{petsc-user-ref}%
  \BibitemOpen
  \bibfield{author}{%
  \bibinfo {author} {\bibfnamefont{S.}~\bibnamefont{Balay}}, \bibinfo {author}
  {\bibfnamefont{K.}~\bibnamefont{Buschelman}}, \bibinfo {author}
  {\bibfnamefont{V.}~\bibnamefont{Eijkhout}}, \bibinfo {author}
  {\bibfnamefont{W.~D.}\ \bibnamefont{Gropp}}, \bibinfo {author}
  {\bibfnamefont{D.}~\bibnamefont{Kaushik}}, \bibinfo {author}
  {\bibfnamefont{M.~G.}\ \bibnamefont{Knepley}}, \bibinfo {author}
  {\bibfnamefont{L.~C.}\ \bibnamefont{McInnes}}, \bibinfo {author}
  {\bibfnamefont{B.~F.}\ \bibnamefont{Smith}},\ and\ \bibinfo {author}
  {\bibfnamefont{H.}~\bibnamefont{Zhang}},\ }%
  \emph{\bibinfo {title} {{PETS}c Users Manual}},\ \bibinfo {type} {Tech.
  Rep.}\ \bibinfo {number} {ANL-95/11 - Revision 3.0.0}\ (\bibinfo
  {institution} {Argonne National Laboratory},\ \bibinfo {year} {2008})%
  \bibAnnoteFile{NoStop}{petsc-user-ref}%
\bibitem{petsc-efficient}%
  \BibitemOpen
  \bibfield{author}{%
  \bibinfo {author} {\bibfnamefont{S.}~\bibnamefont{Balay}}, \bibinfo {author}
  {\bibfnamefont{W.~D.}\ \bibnamefont{Gropp}}, \bibinfo {author}
  {\bibfnamefont{L.~C.}\ \bibnamefont{McInnes}},\ and\ \bibinfo {author}
  {\bibfnamefont{B.~F.}\ \bibnamefont{Smith}},\ }%
  in\ \emph{\bibinfo {booktitle} {Modern Software Tools in Scientific
  Computing}},\ \bibinfo {editor} {edited by\ \bibinfo {editor}
  {\bibfnamefont{E.}~\bibnamefont{Arge}}, \bibinfo {editor}
  {\bibfnamefont{A.~M.}\ \bibnamefont{Bruaset}},\ and\ \bibinfo {editor}
  {\bibfnamefont{H.~P.}\ \bibnamefont{Langtangen}}}\ (\bibinfo {publisher}
  {Birkh{\"{a}}user Press, Boston},\ \bibinfo {year} {1997})\ pp.\ \bibinfo
  {pages} {163--202}%
  \bibAnnoteFile{NoStop}{petsc-efficient}%
\bibitem{Broyden1}%
  \BibitemOpen
  \bibfield{author}{%
  \bibinfo {author} {\bibfnamefont{C.~G.}\ \bibnamefont{Broyden}},\ }%
  \bibfield{journal}{%
  \bibinfo {journal} {J. Inst. Math. Appl.}\ }%
  \textbf{\bibinfo {volume} {6}},\ \bibinfo {pages} {76} (\bibinfo {year}
  {1970})%
  \bibAnnoteFile{NoStop}{Broyden1}%
\bibitem{Fletcher1}%
  \BibitemOpen
  \bibfield{author}{%
  \bibinfo {author} {\bibfnamefont{R.}~\bibnamefont{Fletcher}},\ }%
  \bibfield{journal}{%
  \bibinfo {journal} {Comput. J.}\ }%
  \textbf{\bibinfo {volume} {13}},\ \bibinfo {pages} {317} (\bibinfo {year}
  {1970})%
  \bibAnnoteFile{NoStop}{Fletcher1}%
\bibitem{Goldfarb1}%
  \BibitemOpen
  \bibfield{author}{%
  \bibinfo {author} {\bibfnamefont{D.}~\bibnamefont{Goldfarb}},\ }%
  \bibfield{journal}{%
  \bibinfo {journal} {Math. Comput.}\ }%
  \textbf{\bibinfo {volume} {24}},\ \bibinfo {pages} {23} (\bibinfo {year}
  {1970})%
  \bibAnnoteFile{NoStop}{Goldfarb1}%
\bibitem{Shanno1}%
  \BibitemOpen
  \bibfield{author}{%
  \bibinfo {author} {\bibfnamefont{D.~F.}\ \bibnamefont{Shanno}},\ }%
  \bibfield{journal}{%
  \bibinfo {journal} {Math. Comput.}\ }%
  \textbf{\bibinfo {volume} {24}},\ \bibinfo {pages} {647} (\bibinfo {year}
  {1970})%
  \bibAnnoteFile{NoStop}{Shanno1}%
\bibitem{Poole2}%
  \BibitemOpen
  \bibfield{author}{%
  \bibinfo {author} {\bibfnamefont{C.~P.}\ \bibnamefont{Poole}}, \bibinfo
  {author} {\bibfnamefont{H.}~\bibnamefont{Farach}}, \bibinfo {author}
  {\bibfnamefont{R.}~\bibnamefont{Creswick}},\ and\ \bibinfo {author}
  {\bibfnamefont{R.}~\bibnamefont{Prozorov}},\ }%
  \emph{\bibinfo {title} {Superconductivity, Second Edition}}\ (\bibinfo
  {publisher} {Academic Press},\ \bibinfo {year} {2007})\ \bibinfo
  {pages} {pp.\ 347}%
  \bibAnnoteFile{NoStop}{Poole2}%
\bibitem{Peurla1}%
  \BibitemOpen
  \bibfield{author}{%
  \bibinfo {author} {\bibfnamefont{M.}~\bibnamefont{Peurla}}, \bibinfo {author}
  {\bibfnamefont{H.}~\bibnamefont{Huhtinen}},\ and\ \bibinfo {author}
  {\bibfnamefont{P.}~\bibnamefont{Paturi}},\ }%
  \bibfield{journal}{%
  \bibinfo {journal} {Supercond. Sci. Technol.}\ }%
  \textbf{\bibinfo {volume} {18}},\ \bibinfo {pages} {628} (\bibinfo {year}
  {2005})%
  \bibAnnoteFile{NoStop}{Peurla1}%
\bibitem{Huhtinen12}%
  \BibitemOpen
  \bibfield{author}{%
  \bibinfo {author} {\bibfnamefont{H.}~\bibnamefont{Huhtinen}}, \bibinfo
  {author} {\bibfnamefont{M.}~\bibnamefont{Peurla}}, \bibinfo {author}
  {\bibfnamefont{M.~A.}\ \bibnamefont{Shakhov}}, \bibinfo {author}
  {\bibfnamefont{Y.~P.}\ \bibnamefont{Stepanov}}, \bibinfo {author}
  {\bibfnamefont{P.}~\bibnamefont{Paturi}}, \bibinfo {author}
  {\bibfnamefont{J.}~\bibnamefont{Raittila}}, \bibinfo {author}
  {\bibfnamefont{R.}~\bibnamefont{Palai}},\ and\ \bibinfo {author}
  {\bibfnamefont{R.}~\bibnamefont{Laiho}},\ }%
  \bibfield{journal}{%
  \bibinfo {journal} {IEEE T. Appl. Supercond.}\ }%
  \textbf{\bibinfo {volume} {17}},\ \bibinfo {pages} {3620} (\bibinfo {year}
  {2007})%
  \bibAnnoteFile{NoStop}{Huhtinen12}%
\bibitem{Peurla4}%
  \BibitemOpen
  \bibfield{author}{%
  \bibinfo {author} {\bibfnamefont{M.}~\bibnamefont{Peurla}}, \bibinfo {author}
  {\bibfnamefont{H.}~\bibnamefont{Huhtinen}}, \bibinfo {author}
  {\bibfnamefont{Y.~Y.}\ \bibnamefont{Tse}}, \bibinfo {author}
  {\bibfnamefont{J.}~\bibnamefont{Raittila}},\ and\ \bibinfo {author}
  {\bibfnamefont{P.}~\bibnamefont{Paturi}},\ }%
  \bibfield{journal}{%
  \bibinfo {journal} {IEEE T. Appl. Supercond.}\ }%
  \textbf{\bibinfo {volume} {17}},\ \bibinfo {pages} {3608} (\bibinfo {year}
  {2007})%
  \bibAnnoteFile{NoStop}{Peurla4}%
\bibitem{Huhtinen16}%
  \BibitemOpen
  \bibfield{author}{%
  \bibinfo {author} {\bibfnamefont{H.}~\bibnamefont{Huhtinen}}, \bibinfo
  {author} {\bibfnamefont{K.}~\bibnamefont{Schlesier}},\ and\ \bibinfo {author}
  {\bibfnamefont{P.}~\bibnamefont{Paturi}},\ }%
  \bibfield{journal}{%
  \bibinfo {journal} {Supercond. Sci. Technol.}\ }%
  \textbf{\bibinfo {volume} {22}},\ \bibinfo {pages} {075019} (\bibinfo {year}
  {2009})%
  \bibAnnoteFile{NoStop}{Huhtinen16}%
\bibitem{Huhtinen20}%
  \BibitemOpen
  \bibfield{author}{%
  \bibinfo {author} {\bibfnamefont{H.}~\bibnamefont{Huhtinen}}, \bibinfo
  {author} {\bibfnamefont{M.}~\bibnamefont{Irjala}}, \bibinfo {author}
  {\bibfnamefont{P.}~\bibnamefont{Paturi}},\ and\ \bibinfo {author}
  {\bibfnamefont{M.}~\bibnamefont{Falter}},\ }%
  \bibfield{journal}{%
  \bibinfo {journal} {IEEE T. Appl. Supercond.}\ }%
  \textbf{\bibinfo {volume} {21}},\ \bibinfo {pages} {2753} (\bibinfo {year}
  {2011})%
  \bibAnnoteFile{NoStop}{Huhtinen20}%
\bibitem{Bezryadin2}%
  \BibitemOpen
  \bibfield{author}{%
  \bibinfo {author} {\bibfnamefont{A.}~\bibnamefont{Bezryadin}}, \bibinfo
  {author} {\bibfnamefont{Y.~N.}\ \bibnamefont{Ovchinnikov}},\ and\ \bibinfo
  {author} {\bibfnamefont{B.}~\bibnamefont{Pannetier}},\ }%
  \bibfield{journal}{%
  \bibinfo {journal} {Phys. Rev. B}\ }%
  \textbf{\bibinfo {volume} {53}},\ \bibinfo {pages} {8553} (\bibinfo {year}
  {1996})%
  \bibAnnoteFile{NoStop}{Bezryadin2}%
\end{thebibliography}

%

\end{document}